\documentclass[a4paper,12pt]{article}

\usepackage{graphicx}
\usepackage{amsmath}
\usepackage{amssymb}
\usepackage{hyperref}
\usepackage{tabularx}
\newcolumntype{M}{>{$}c<{$}}
\usepackage[matrix,arrow,curve]{xy}

\numberwithin{equation}{section} \numberwithin{figure}{section}
\numberwithin{table}{section}

\def\papertitlepage{\baselineskip 3.5ex\thispagestyle{empty}}
\def\Title#1{\baselineskip 1cm \vspace{1.5cm}%
  \begin{center}{\Large\bf #1}\end{center}\vspace{0.5cm}}
\def\Authors#1{\begin{center}\renewcommand{\thefootnote}{\fnsymbol{footnote}}{\it #1}\end{center}}
\def\Abstract{\vspace{1.0cm}%
  \begin{center}{\large\bf Abstract}\end{center}}

\renewenvironment{thebibliography}{\pagebreak[3]\par\vspace{0.6em}
\begin{flushleft}{\large \bf References}\end{flushleft}
\vspace{-1.0em}

\begin{enumerate}\if@twocolumn\baselineskip=0.6em\itemsep -0.2em
\else\itemsep -0.2em\fi\labelsep 0.1em}{\end{enumerate} }

\DeclareMathDelimiter{\lcolon}{\mathopen}{operators}{"3A}{largesymbols}{"3A}
\DeclareMathDelimiter{\rcolon}{\mathclose}{operators}{"3A}{largesymbols}{"3A}

\def\+{\!\!+\!\!}

\def\dynkin(#1){(#1)}

\def\bra<#1|{\langle#1|}
\def\ket|#1>{|#1\rangle}
\def\braket<#1|#2>{\langle#1|#2\rangle}
\def\llangle{\langle\!\langle}
\def\rrangle{\rangle\!\rangle}
\def\bbra<#1|{\llangle#1|}
\def\kket|#1>{|#1\rrangle}
\def\bbraket<#1|#2>{\llangle#1|#2\rrangle}

\begin{document}
{\papertitlepage \vspace*{0cm} {\hfill
\begin{minipage}{4.2cm}
IFT-P. 2010\par\noindent August, 2010
\end{minipage}}
\Title{Generating Erler-Schnabl-type Solution for Tachyon Vacuum
in Cubic Superstring Field Theory}
\Authors{{\sc E.~Aldo~Arroyo${}$\footnote{\tt
aldohep@ift.unesp.br}}
\\
Instituto de F\'{i}sica Te\'{o}rica, UNESP - Universidade Estadual Paulista \\[-2ex]
Caixa Postal 70532-2 \\ 01156-970 S\~{a}o Paulo, SP, Brasil
\\
${}$ }
} 

\vskip-\baselineskip
{\baselineskip .5cm \Abstract We study a new set of identity-based
solutions to analyze the problem of tachyon condensation in open
bosonic string field theory and cubic superstring field theory.
Even though these identity-based solutions seem to be trivial, it
turns out that after performing a suitable gauge transformation,
we are left with the known Erler-Schnabl-type solutions which
correctly reproduce the value of the D-brane tension. This result
shows explicitly that how a seemingly trivial solution can
generate a non-trivial configuration which precisely represents to
the tachyon vacuum.
 }
\newpage
\setcounter{footnote}{0}
\tableofcontents

\section{Introduction}
Recently it was found a simple analytic solution for tachyon
condensation in open bosonic string field theory
\cite{Erler:2009uj} as well as in the modified cubic superstring
field theory \cite{Gorbachev:2010zz}. The main feature of these
solutions is that, rather than a discrete sum, the solutions
(which we refer as the Erler-Schnabl-type solutions) can be
written as a continuous integral over wedge states, where no
regularization or phantom term is required. Since these solutions
do not require the presence of the phantom term, in contrast to
the $\mathcal{B}_0$ gauge solutions
\cite{Schnabl:2005gv,Okawa:2006vm,Fuchs:2006hw,Ellwood:2006ba,
Arroyo:2009ec,AldoArroyo:2009hf,Erler:2006hw,Erler:2006ww,Erler:2007xt,Aref'eva:2009ac,Aref'eva:2009sj,Aref'eva:2008ad},
the computation of the value of the D-brane tension is very
straightforward.

It would be interesting to find a similar Erler-Schnabl-type
solution in the case of Berkovits WZW-type superstring field
theory \cite{Berkovits:1995ab}. Since the action for this theory
is a non-polynomial one, the issue for finding the tachyon vacuum
solution and the computation of the value of the D-brane tension
seems to be highly cumbersome. Therefore, in order to find some
clues for this challenging puzzle, we should analyze the problem
in a relatively simple formulation of open superstring field
theory.

In the literature, there is an old formulation of open superstring
field theory, namely the modified cubic superstring field theory
\cite{Preitschopf:1989fc,Arefeva:1989cp}. The action in this
formulation is cubic (instead of the non-polynomial action given
by Berkovits \cite{Berkovits:1995ab})
\begin{eqnarray}
\label{accionI1} S=-\frac{1}{g^2}\Big[ \frac{1}{2} \langle \langle
\Phi,Q \Phi \rangle \rangle +\frac{1}{3} \langle \langle
\Phi,\Phi*\Phi \rangle \rangle \Big] \, ,
\end{eqnarray}
where $Q$ is the BRST operator, $*$ stands for Witten's star
product \cite{Witten:1985cc}, and the inner product $\langle
\langle \cdot , \cdot \rangle \rangle$ is the standard BPZ inner
product with the difference that we must insert the operator
$Y_{-2}$ at the open string midpoint. The operator $Y_{-2}$ can be
written as the product of two inverse picture changing operators
$Y_{-2}=Y(i)Y(-i)$, where $Y(z)=-\partial \xi e^{-2 \phi} c(z)$.
The string field $\Phi$ which has ghost number 1 and picture
number 0 belongs to the small Hilbert space of the first-quantized
matter+ghost open Neveu-Schwarz superstring theory.

As stated in Sen's first conjecture
\cite{Sen:1999mh,Sen:1999xm,Bagchi:2008et}, the classical open
string field equation of motion should admit a Poincar\'{e}
invariant solution $\Phi\equiv\Psi$ which is identified as the
tachyon vacuum with no D-branes. This statement means that the
energy density of the true vacuum found by solving the equation of
motion should be equal to minus the tension of the D-brane. Since
the energy density of a static configuration is minus the action,
for the case of the cubic action, Sen's conjecture can be
summarized as follows
\begin{eqnarray}
\label{poten1} \frac{1}{g^2}\Big[\frac{1}{2} \langle \langle
\Psi,Q \Psi \rangle \rangle +\frac{1}{3} \langle \langle
\Psi,\Psi*\Psi \rangle \rangle \Big]= - \frac{1}{2 \pi^2 g^2} \, .
\end{eqnarray}
The string field equation of motion and Sen's conjecture allow us
to fix the kinetic term,
\begin{eqnarray}
\label{kinec1} \frac{\pi^2}{3} \langle \langle \Psi,Q \Psi \rangle
\rangle &=&-1 \, .
\end{eqnarray}

In this paper, we propose a new prescription for generating a
string field $\Psi$, in cubic string field theories
\cite{Preitschopf:1989fc,Arefeva:1989cp,Witten:1985cc}, which
satisfies the equation of motion $Q \Psi + \Psi*\Psi =0$ and
represents to the tachyon vacuum. The procedure of our
prescription follows two steps: (i) find a naive identity-based
solution \cite{Kishimoto:2009nd} of the string field equation of
motion, and (ii) perform a gauge transformation
\cite{Ellwood:2009zf} over the identity-based solution such that
the resulting string field, consistently, represents to the
tachyon vacuum\footnote{By consistently we mean that the solution
must reproduce correctly and unambiguously the value of the
D-brane tension.}.

In order to find identity-based solutions\footnote{By
identity-based solution we mean a solution which is based on the
identity string field $\mathcal{I}$.} to the string field
equations of motion, we are going to use a basis similar to the
one used in \cite{Schnabl:2005gv} with the difference that in our
case the operators are inserted on the identity string field
$\mathcal{I}$\footnote{Let us remember that the identity string
field is the wedge state $\mathcal{I}=U_{1}^\dag U_{1} |0\rangle$.
A discussion, needed for the purpose of this paper, about the
identity string field can be found in the reference
\cite{Schnabl:2002gg}.}. For instance in the case of open bosonic
string field theory, to find a solution based on the identity
string field, we should use the ansatz
\begin{eqnarray}
\label{ansa01} \Psi=\sum_{n,p} f_{n,p}U_{1}^\dag U_{1}
\hat{\mathcal{L}}^n \tilde{c}_p |0\rangle +\sum_{n,p,q} f_{n,p,q}
U_{1}^\dag U_{1} \hat{\mathcal{B}}\hat{\mathcal{L}}^n \tilde{c}_p
\tilde{c}_q |0\rangle \, ,
\end{eqnarray}
where $n = 0, 1, 2, \cdots$, and $p, q = 1, 0,-1,-2, \cdots \cdot$
The operators $\hat{\mathcal{L}}^n$, $\hat{\mathcal{B}}$ and
$\tilde{c}_p$ are defined in \cite{Schnabl:2005gv}. Plugging this
ansatz (\ref{ansa01}) into the equation of motion will lead to a
system of algebraic equations for the coefficients $f_{n,p}$ and
$f_{n,p,q}$. Analyzing these algebraic equations we discover that
many of the coefficients can be set to zero, therefore we can use
a simpler ansatz than the one given by (\ref{ansa01}), for
instance in this paper we are going to use the following ansatz
\begin{eqnarray}
\label{ansa002} \Psi = \alpha_1 c + \alpha_2 cK + \alpha_3 Kc \, ,
\end{eqnarray}
where the basic string fields $K$ and $c$ together with $B$ are
defined in terms of the identity string field with the
appropriated insertions \cite{Erler:2006hw,Erler:2006ww}
\begin{eqnarray}
\label{K01in} K \equiv \frac{1}{\pi} \hat{\mathcal{L}} U_{1}^\dag
U_{1} |0\rangle \, , \;\;\; c\equiv   U_{1}^\dag U_{1} \tilde
c_1|0\rangle \, , \;\;\; B \equiv \frac{1}{\pi} \hat{\mathcal{B}}
U_{1}^\dag U_{1} |0\rangle \, .
\end{eqnarray}

As it will be shown, the solution obtained from the ansatz
(\ref{ansa002}) provides ambiguous analytic result for the value
of the vacuum energy and consequently for the D-brane tension. In
general, it can be shown that the energy of and identity-based
solution is of the form $0 \times \infty$ \cite{Kishimoto:2009nd},
and so it is not well defined. Instead of trying to find a
consistent regularization in order to treat our identity-based
solution correctly, we are going to show that this particular
identity-based solution is equivalent to the recently analytic
solution found by Erler and Schnabl \cite{Erler:2009uj}.

To show the above statement, we need to find an explicit gauge
transformation which relates the identity-based solution with the
Erler-Schnabl solution. The explicit form of this gauge
transformation is given by
\begin{eqnarray}
\label{regularsol1} \Psi_{\text{E-S}} = U Q U^{-1} + U \Psi_I
U^{-1} \, ,
\end{eqnarray}
where $\Psi_{\text{E-S}}$ is the Erler-Schnabl solution
\cite{Erler:2009uj}, $\Psi_I$ is the identity-based solution, and
$U$ is an element of the gauge transformation
\begin{eqnarray}
\label{gauge01} U=1+ cBK\, , \;\;\; U^{-1}=1- cBK \frac{1}{1+ K}
\, .
\end{eqnarray}
Since the Erler-Schnabl solution contains a term depending on
$1/(1+K)$, which is crucial for the computation of the vacuum
energy \cite{Erler:2009uj}, it is not so difficult to guess the
form of the gauge transformation\footnote{Let us remark that in
general we can write expressions for $U$ and $U_{-1}$ depending on
a function of $K$ \begin{eqnarray}  U=1+ cB f(K)\, , \;\;\;
U^{-1}=1- cB \frac{f(K)}{1+ f(K)} \nonumber \, .
\end{eqnarray}} which relates our identity-based solution with Erler-Schnabl
solution, and in particular the gauge transformation should
contain the $1/(1+K)$ term.

Nevertheless we can construct different tachyon vacuum solutions
by replacing the term $1/(1+K)$ with other functions of $K$
satisfying the criteria given in reference \cite{Erler:2006ww}.
For example the dependence on $K$ in the original Schnabl's
solution \cite{Schnabl:2005gv} is given by $-e^{-K}$
\cite{Okawa:2006vm,Schnabl:2010tb}. It is clear that if we want to
relate our identity-based solution to these different tachyon
vacuum solutions we should use another gauge transformation which
will depend on the choice of the function of $K$. The reason why
we chose the $1/(1+K)$ dependence is because the computation of
the vacuum energy is simplified. It is well known that the case
$-e^{-K}$ which corresponds to the original Schnabl's solution has
subtleties \cite{Schnabl:2005gv,Okawa:2006vm,Schnabl:2010tb}, for
instance in order to compute the right value of the vacuum energy
the phantom term must be included.

Finally, we carry out the same analysis for the case of the
modified cubic superstring field theory, namely we show that a
solution based on the identity string field can be brought to the
tachyon vacuum solution constructed by Gorbachev using a gauge
transformation. Our results show explicitly that how a seemingly
trivial identity-based solution can generate a non-trivial
configuration which precisely represents to the tachyon vacuum.
Certainly it would be very interesting to extend our results to
the case of the non-polynomial Berkovits WZW-type superstring
field theory.

This paper is organized as follows. In section 2, we review and
further develop some properties of the simple analytic solution
for tachyon condensation in open bosonic string field theory. We
give an example of an identity-based solution which formally
solves the equation of motion. After performing a gauge
transformation over this seemingly trivial identity-based
solution, we obtain Erler-Schnabl's solution which correctly
reproduces the value of the D-brane tension. We also show that
Erler-Schnabl's solution appears as a particular case of a rather
general two-parameter family of solutions. In section 3, we
analyze a similar identity-based solution in the modified cubic
superstring field theory. As in the bosonic case, by performing a
suitable gauge transformation over this identity-based solution,
we obtain the known Gorbachev's solution which correctly
reproduces the value of the vacuum energy. A two-parameter family
of solutions is discussed as well. In section 4, a summary and
further directions of exploration are given. The details related
to the explicit construction of identity-based solutions and the
gauge transformation are left for the appendix.

\section{Simple analytic solution in the bosonic case}

In this section, after review some aspects of the simple analytic
solution for tachyon condensation in open bosonic string field
theory \cite{Erler:2009uj}, we analyze a new identity-based
solution which formally solves the equation of motion. Since this
solution is an isolated identity-like piece, it may happen that
the solution is trivial or inconsistent, in fact similar
identity-based solutions have been proposed in the past, and for
such solutions there is no unambiguous analytic calculation for
the D-brane tension \cite{Kishimoto:2009nd}. However, as we are
going to see, after performing a gauge transformation over this
seemingly trivial identity-based solution, we obtain the well
known Erler-Schnabl's solution which correctly reproduces the
value of the D-brane tension. We also show that Erler-Schnabl's
solution appears as a particular case of a rather general
two-parameter family of solutions.

\subsection{Erler-Schnabl's solution}
As derived in Erler-Schnabl's paper \cite{Erler:2009uj} using the
methods of \cite{Schnabl:2005gv,Erler:2006ww}, the simple analytic
solution for tachyon condensation in open bosonic string field
theory is
\begin{eqnarray}
\label{ESsolution} \Psi_{\text{E-S}} = (c + cKBc)\frac{1}{1+K} \,
,
\end{eqnarray}
where the basic string fields $K$, $B$ and $c$ are given in the
split string notation
\cite{Okawa:2006vm,Erler:2006hw,Erler:2006ww}, and they can be
written, using the operator representation \cite{Schnabl:2005gv},
as follows
\begin{eqnarray}
\label{K01} K &\rightarrow& \frac{1}{\pi} \hat{\mathcal{L}}
U_{1}^\dag U_{1} |0\rangle \, ,
\\
\label{B01} B &\rightarrow& \frac{1}{\pi} \hat{\mathcal{B}}
U_{1}^\dag U_{1} |0\rangle \, ,
\\
\label{c01} c &\rightarrow&   U_{1}^\dag U_{1} \tilde c
(0)|0\rangle \, .
\end{eqnarray}

The operators $\hat{\mathcal{L}}$, $\hat{\mathcal{B}}$ and $\tilde
c(0)$ are defined in the sliver frame \cite{AldoArroyo:2009hf}
\footnote{Remember that a point in the upper half plane $z$ is
mapped to a point in the sliver frame $\tilde z$ via the conformal
mapping $\tilde z= \arctan z $.}, and they are related to the
worldsheet energy-momentum tensor, the $b$ and $c$ ghosts fields
respectively, for instance
\begin{eqnarray}
\label{Lhat01} \hat{\mathcal{L}} &\equiv& \mathcal{L}_{0} +
\mathcal{L}^{\dag}_0 = \oint \frac{d z}{2 \pi i} (1+z^{2})
(\arctan z+\text{arccot} z) \,
T(z) \, , \\
\label{Bhat01} \hat{\mathcal{B}} &\equiv& \mathcal{B}_{0} +
\mathcal{B}^{\dag}_0 = \oint \frac{d z}{2 \pi i} (1+z^{2})
(\arctan z+\text{arccot} z) \, b(z) \, ,
\end{eqnarray}
while the operator $U_{1}^\dag U_{1}$ in general is given by
$U^\dag_r U_r = e^{\frac{2-r}{2} \hat{\mathcal{L}}}$, so we have
chosen $r=1$, note that the string field $U_{1}^\dag U_{1}
|0\rangle$ represents to the identity string field $1 \rightarrow
U_{1}^\dag U_{1} |0\rangle$
\cite{Schnabl:2005gv,Okawa:2006vm,Erler:2006hw,Erler:2006ww}.

Using the operator representation (\ref{K01})-(\ref{c01}) of the
string fields $K$, $B$ and $c$, we can show that these fields
satisfy the algebraic relations
\begin{eqnarray}
\label{eq2} \{B,c\}=1\, , \;\;\;\;\;\;\; [B,K]=0 \, ,
\;\;\;\;\;\;\; B^2=c^2=0 \, ,
\end{eqnarray}
and have the following BRST variations
\begin{eqnarray}
\label{eq3} QK=0 \, , \;\;\;\;\;\; QB=K \, , \;\;\;\;\;\; Qc=cKc
\, .
\end{eqnarray}

As we can see, the solution (\ref{ESsolution}) contains explicitly
to the identity string field and therefore, we may think that this
solution is not well defined in the sense that for such solution
there is no unambiguous analytic calculation for the D-brane
tension. However, as shown in \cite{Erler:2009uj}, Erler-Schnabl's
solution (\ref{ESsolution}) correctly and unambiguously reproduces
the value of the D-brane tension. To prove this statement is
crucial to write the solution as the following integral
\begin{eqnarray}
\label{ISol01} \Psi_{\text{E-S}} = \int_{0}^{\infty} dt \,
e^{-t}(c + cKBc) \Omega^t \, ,
\end{eqnarray}
this form of the solution is possible since we can invert $1+K$
using the Schwinger parameterization
\begin{eqnarray}
\label{1K01} \frac{1}{1+K} = \int_{0}^{\infty} dt \, e^{-t(1+K)} =
\int_{0}^{\infty} dt \, e^{-t} \Omega^t \, .
\end{eqnarray}
As stated in \cite{Erler:2009uj}, the fact that the solution can
be written in terms of a continuous integral over wedge states
arbitrarily close to the identity, and not as an isolated
identity-like piece, is crucial for the consistency of the
solution.

\subsection{The identity-based solution and gauge transformation}
In this subsection, we are going to analyze a new simple
identity-based solution to the equation of motion in open bosonic
string field theory, this solution is given by
\begin{eqnarray}
\label{eq1} \Psi=c(1-K) \, .
\end{eqnarray}
Using the algebraic relations (\ref{eq2}) and (\ref{eq3}), it is
easy to show that the string field $\Psi$ (\ref{eq1}) satisfies
the equation of motion
\begin{eqnarray}
\label{eq5} Q \Psi + \Psi^2 =0 \, .
\end{eqnarray}

The next step is to compute the value of the vacuum energy, this
computation is crucial if we want to verify Sen's first
conjecture. Using the equation of motion, the computation of the
energy can be reduced to the evaluation of the following
correlator
\begin{eqnarray}
\label{eq7} \frac{1}{6 g^2} \langle \Psi Q \Psi \rangle &=&
\frac{1}{6 g^2} \langle c(1-K)Q(c(1-K))
\rangle \nonumber \\
&=&\frac{1}{6 g^2} \big[\langle c^2 K c \rangle - \langle cKcKc
\rangle + \langle
c^2KcK \rangle + \langle cKcKcK \rangle  \big] \nonumber \\
&=& 0 \, .
\end{eqnarray}
The first term and third term vanish because $c^2=0$, while the
second vanishes for the same reason upon using cyclicity of the
correlator. The las term
\begin{eqnarray}
 cKcKcK = cKc \partial c K = c (\partial c)^2 K
\end{eqnarray}
vanishes because $(\partial c)^2$. Based on this computation we
conclude that the identity-based solution (\ref{eq1}) seems to be
trivial.

Though we have a vanishing result for the value of the vacuum
energy, it would be desirable to confirm our calculation by other
means, for instance using the $\mathcal{L}_{0}$-level expansion of
the identity-based solution (\ref{eq1})
\begin{eqnarray}
\label{L0identity01} \Psi&=&  U_{1}^\dag U_{1} \big[  \tilde c_1 -
\frac{1}{\pi} \hat{\mathcal{L}} \tilde c_1 + \frac{1}{2}
 \tilde c_0 \big] |0\rangle  \nonumber \\
&=&\sum_{n=0}^{\infty} \big[ \frac{1}{2^n n!} \hat{\mathcal{L}}^n
\tilde c_1 - \frac{1}{\pi} \frac{1}{2^n n!}
\hat{\mathcal{L}}^{n+1} \tilde c_1  + \frac{1}{2^{n+1} n!}
\hat{\mathcal{L}}^n \tilde c_0 \big] |0\rangle  \nonumber \\
&=&\sum_{n=0}^{\infty}\sum_{p=0}^{1} f_{n,p} \hat{\mathcal{L}}^n
\tilde c_p |0\rangle \, ,
\end{eqnarray}
where the coefficient $f_{n,p}$ is given by
\begin{align}
f_{n,p} =
\begin{cases}
 \frac{1}{2^{n+1} n!} \; ,  & \mbox{if }n \geq 0 \; \mbox{and }p = 0 \\
  1,  & \mbox{if }n = 0 \; \mbox{and }p = 1\\
 \frac{1}{2^n
n!}-\frac{1}{\pi} \frac{1}{2^{n-1} (n-1)!} \, , & \mbox{if }n \geq
1 \; \mbox{and }p = 1 \; .
\end{cases}
\end{align}
Using the $\mathcal{L}_{0}$-level expansion of our identity-based
solution (\ref{L0identity01}), we find that the value of the
vacuum energy is given by
\begin{eqnarray}
\label{L0energy01} \frac{1}{6 g^2} \langle \Psi Q \Psi \rangle &=&
\frac{1}{6 g^2} \sum_{m=0}^{\infty}\sum_{p=0}^{1}
\sum_{n=0}^{\infty}\sum_{q=0}^{1} f_{m,p}f_{n,q}\langle
\text{bpz}(\tilde{c}_{p}) \hat{\mathcal{L}}^{m+n} Q \tilde{c}_{q}
\rangle \nonumber \\
&=& - \frac{1}{6 g^2} \big[ \frac{1}{2} + \frac{2}{\pi^2} \big] \,
.
\end{eqnarray}

Therefore, as we can see, this solution provides ambiguous
analytic result for the value of the vacuum energy and
consequently for the D-brane tension. It is intriguing to note
that even though we have obtained a value different from zero
(\ref{L0energy01}), this value does not coincide with the one
given in equation (\ref{kinec1}). To understand this anomaly
better, let us consider two general identity-like string fields
with $\mathcal{L}_{0}$ eigenvalues equal to $h_1$ and $h_2$
respectively
\begin{eqnarray}
\label{identitySF1} \tilde \phi_1 = U_{1}^\dag U_{1} \tilde \phi_1
(0)|0\rangle \, , \;\;\;\; \tilde \phi_2 = U_{1}^\dag U_{1} \tilde
\phi_2 (0)|0\rangle \, ,
\end{eqnarray}
and compute the correlator $\langle \tilde \phi_1, \tilde \phi_2
\rangle$
\begin{eqnarray}
\label{correau01} \langle \tilde \phi_1, \tilde \phi_2 \rangle &=&
\langle 0| \text{bpz}\big(\tilde{\phi}_{1}(0)\big) U_{1}^\dag
U_{1} U_{1}^\dag U_{1} \tilde \phi_2 (0) |0\rangle
\nonumber \\
 &=&
\lim_{r\rightarrow 1}\langle 0|
\text{bpz}\big(\tilde{\phi}_{1}(0)\big) U_{r}^\dag U_{r}
U_{r}^\dag U_{r} \tilde \phi_2 (0) |0\rangle \nonumber \\
&=& \lim_{r\rightarrow 1} \big(\frac{2}{r}\big)^{h_1+h_2}\langle
0| \text{bpz}\big(\tilde{\phi}_{1}(0)\big)  U_{r} U_{r}^\dag
\tilde \phi_2 (0) |0\rangle \nonumber \\
&=& \lim_{r\rightarrow 1} \Big[ \Big(\frac{1}{r-1}\Big)^{h_1+h_2}
\Big]\langle 0| \text{bpz}\big(\tilde{\phi}_{1}(0)\big)  \tilde
\phi_2 (0) |0\rangle \, ,
\end{eqnarray}
where we have used the definition of
$U_r=\big(\frac{2}{r}\big)^{\mathcal{L}_{0}}$. Clearly if the sum
of the eigenvalues $h_1 + h_2$ is greater than zero the correlator
is divergent, while in the case $h_1 + h_2=0$ the correlator is
ambiguous. Only when the sum is less than zero we obtain an
unambiguous result $\langle \tilde \phi_1, \tilde \phi_2
\rangle=0$. Since the case $h_1 + h_2>0$ could appear in the
computation of the vacuum energy, potential singularities can be
present. This analysis also shows the origin of the non-triviality
of the identity based solution.

In this paper, instead of trying to find a consistent
regularization in order to treat our identity-based solution
correctly, we are going to show that this particular
identity-based solution (\ref{eq1}) is equivalent to the recently
analytic solution found by Erler and Schnabl, to show this
statement we need to find an explicit gauge transformation which
relates the identity-based solution (\ref{eq1}) with
(\ref{ESsolution}), and in fact we have found the explicit form of
this gauge transformation
\begin{eqnarray}
\Psi &=& U^{-1} (Q + \Psi_{\text{E-S}}) U \nonumber \\
&=& \Big[1-cBK\frac{1}{1+K}\Big] \Big( Q + (c + cKBc)\frac{1}{1+K}
\Big) \Big[cBK +1\Big] \nonumber \\
&=&\Big[1-cBK\frac{1}{1+K}\Big] \Big( Q(cKB+1) +
(c+cKBc)\frac{1}{1+K}(cBK+1) \Big) \nonumber \\
&=& \Big[1-cBK\frac{1}{1+K}\Big] \Big(  cKcKB -cK^2 + c
\frac{1}{1+K} - cK \frac{1}{1+K}K + cKBc \Big) \nonumber \\
&=& \Big[1-cBK\frac{1}{1+K}\Big] \Big(  cKcKB -cK^2 + c
\frac{1}{1+K}(1-K^2) + cKBc \Big) \nonumber \\
&=&\Big[1-cBK\frac{1}{1+K}\Big] \Big(  cKcKB -cK^2 + c(1-K) + cKBc
\Big) \nonumber \\
&=&\Big[1-cBK\frac{1}{1+K}\Big] (cBK+1) \Big( c(1-K) \Big)
\nonumber \\
&=& c (1-K) \, .
\end{eqnarray}

So we just have shown, by explicit computations, that starting
with a seemingly trivial identity-based solution (\ref{eq1}), by a
suitable gauge transformation, we arrive to a well behaved
solution which correctly reproduce the desire value of the D-brane
tension. In the next subsection, we are going to show that the
Erler-Schnabl's solution appears as a particular case of a rather
general two-parameter family of solutions.

\subsection{Two-parameter family of solutions}
As it was mentioned in the introduction section, in order to find
identity-based solutions to the open bosonic string field
equations of motion, we should use the following ansatz
\begin{eqnarray}
\label{ansa01subsec} \Psi=\sum_{n,p} f_{n,p}U_{1}^\dag U_{1}
\hat{\mathcal{L}}^n \tilde{c}_p |0\rangle +\sum_{n,p,q} f_{n,p,q}
U_{1}^\dag U_{1} \hat{\mathcal{B}}\hat{\mathcal{L}}^n \tilde{c}_p
\tilde{c}_q |0\rangle \, ,
\end{eqnarray}
where $n = 0, 1, 2, \cdots$, and $p, q = 1, 0,-1,-2, \cdots \cdot$
The operators $\hat{\mathcal{L}}^n$, $\hat{\mathcal{B}}$ and
$\tilde{c}_p$ are defined in \cite{Schnabl:2005gv}. Plugging this
ansatz (\ref{ansa01subsec}) into the equation of motion will lead
to a system of algebraic equations for the coefficients $f_{n,p}$
and $f_{n,p,q}$. Analyzing these algebraic equations we discover
that many of the coefficients can be set to zero, therefore we can
use a simpler ansatz than the one given by (\ref{ansa01subsec}),
for instance we are going to use the following ansatz
\begin{eqnarray}
\label{ansa002sub} \Psi = \alpha_1 c + \alpha_2 cK + \alpha_3 Kc
\, ,
\end{eqnarray}
Plugging this ansatz (\ref{ansa002sub}) into the equation of
motion $Q \Psi + \Psi \Psi=0$, we obtain an algebraic equation for
the coefficients $\alpha_i$
\begin{eqnarray}
\label{ape02sub} 1+\alpha_2+\alpha_3 =0 \, ,
\end{eqnarray}
and consequently our ansatz (\ref{ansa002sub}) becomes
\begin{eqnarray}
\label{ape03sub} \Psi = \alpha_1 c + \alpha_2 cK - (1+\alpha_2) Kc
\, .
\end{eqnarray}

A string field $\Psi'$ which identically satisfies the string
field equation of motion can be derived by performing a gauge
transformation over the identity-based solution (\ref{ape03sub})
\begin{eqnarray}
\label{ape06sub} \Psi'=U (\Psi+Q) U^{-1} \, .
\end{eqnarray}
Plugging the expressions (\ref{gauge01}) for the string field $U$
and its inverse $U^{-1}$ into the definition of the gauge
transformation (\ref{ape06sub}), we obtain a two-parameter family
of solutions
\begin{eqnarray}
\label{ape07au01} \Psi'= \big[ \alpha_1 (c+ cBKc) + (\alpha_1 +
\alpha_2)( cK + cBKcK) -(1+\alpha_2)( Kc + KcBKc )  \big]
\frac{1}{1+K} \, . \nonumber \\
\end{eqnarray}

As it is shown in the appendix, to simplify the calculation of the
vacuum energy, we can fix the values of the two parameters
$\alpha_1$ and $\alpha_2$. For instance the particular values of
the parameters $\alpha_1=1$ and $\alpha_2=-1$ correspond to the
Erler-Schnabl's solution (\ref{ESsolution}). Nevertheless at this
point it is interesting to ask: do the solutions with different
values of the parameters describe the tachyon vacuum? To provide
an answer to this question, we should compute the vacuum energy
for the solution (\ref{ape07au01}) with arbitrary values for the
parameters $\alpha_1$ and $\alpha_2$.

In order to perform this computation, let us write the solution
(\ref{ape07au01}) as an expression containing an exact BRST term
\begin{align}
\label{correxz01} \Psi' =& \big[ \alpha_1 c + (\alpha_1 +
\alpha_2)cK  -(1+\alpha_2)Kc   \big] \frac{1}{1+ K} \nonumber \\
&+ Q \Big\{ \big[ \alpha_1 Bc + (\alpha_1 + \alpha_2)BcK
-(1+\alpha_2) KBc \big] \frac{1}{1+K} \Big\} \,.
\end{align}
It turns out that the computation of the vacuum energy can be
reduced to the evaluation of the following correlator
\begin{align}
\label{correxz02} \frac{1}{6 g^2}\langle  \Psi',Q \Psi' \rangle =
\frac{1}{6 g^2} \langle ( a_1 c + a_2 cK +a_3Kc  ) \frac{1}{1+ K}
( a_1 cKc + a_2 cKcK +a_3KcKc  ) \frac{1}{1+ K} \rangle \, ,
\end{align}
where we have defined the coefficients
\begin{eqnarray}
\label{correxz03} a_1=\alpha_1 \, , \;\;\;\;\;\; a_2=\alpha_1 +
\alpha_2 \, , \;\;\;\;\;\; a_3=-1-\alpha_2 \, .
\end{eqnarray}

Expanding the right hand side of equation (\ref{correxz02}), we
obtain the following expression for the vacuum energy
\begin{align}
\label{correxz04} &\frac{1}{6 g^2} \big[ a_1^2 \langle  c
\frac{1}{1+ K} cKc \frac{1}{1+ K} \rangle + a_1 a_2 \langle  c
\frac{1}{1+ K} cKcK \frac{1}{1+ K} \rangle +   a_1 a_3 \langle  c
\frac{1}{1+ K} KcKc
\frac{1}{1+ K} \rangle  \nonumber \\
&+a_2a_1 \langle  cK \frac{1}{1+ K} cKc \frac{1}{1+ K} \rangle +
a_2^2 \langle  cK \frac{1}{1+ K} cKcK \frac{1}{1+ K} \rangle + a_2
a_3 \langle  cK \frac{1}{1+ K} KcKc \frac{1}{1+ K} \rangle
\nonumber \\
&+a_3a_1 \langle  Kc \frac{1}{1+ K} cKc \frac{1}{1+ K} \rangle +
a_3a_2 \langle  Kc \frac{1}{1+ K} cKcK \frac{1}{1+ K} \rangle +
a_3^2 \langle  Kc \frac{1}{1+ K} KcKc \frac{1}{1+ K} \rangle
\big].
\end{align}
All the correlators involved in the evaluation of the vacuum
energy (\ref{correxz04}) can be computed using the basic
correlator $\langle  \Omega^{r_1} c \Omega^{r_2} c\Omega^{r_3}c
\Omega^{r_4} \rangle$
\begin{align}
\label{correxz05} \langle  \Omega^{r_1} c \Omega^{r_2}
c\Omega^{r_3}c \Omega^{r_4} \rangle= \frac{L^3}{\pi ^3}\sin
[\frac{\pi r_2}{L}] \sin [\frac{\pi
   r_3}{L}] \sin [\frac{\pi
   (r_2+r_3)}{L}] \, , \;\;\;\;\;\;\; L=r_1+r_2+r_3+r_4 \, .
\end{align}
For instance, the expression for the correlator $\langle Kc
\frac{1}{1+ K} KcKc \frac{1}{1+ K} \rangle$ is given by
\begin{align}
\label{correxz06} \langle Kc \frac{1}{1+ K} KcKc \frac{1}{1+ K}
\rangle &= - \int_{0}^{\infty}\int_{0}^{\infty}dt_1dt_2
e^{-t_1-t_2} \frac{\partial^3 \langle \Omega^{s_1} c
\Omega^{t_1+s_2} c\Omega^{s_3}c \Omega^{t_2} \rangle}{\partial s_1
\partial s_2 \partial s_3}
\Big|_{s_1=0,s_2=0,s_3=0} \nonumber \\
&=-\frac{3}{\pi ^2} \, .
\end{align}

Performing similar computations for the rest of the correlators,
from (\ref{correxz04}) we get the following expression for the
vacuum energy (\ref{correxz02})
\begin{align}
\label{correxz07} \frac{1}{6 g^2}\langle  \Psi',Q \Psi' \rangle =
-\frac{1}{2 \pi ^2 g^2 }(-a_1+a_2+a_3)^2 \, ,
\end{align}
plugging the definition of the coefficients $a_1$, $a_2$, $a_3$
(\ref{correxz03}) into the equation (\ref{correxz07}), we
remarkably obtain the right value for the vacuum energy
\begin{align}
\label{correxz08} \frac{1}{6 g^2}\langle  \Psi',Q \Psi' \rangle =
-\frac{1}{2 \pi ^2 g^2 } \, ,
\end{align}
therefore this last result (\ref{correxz08}) shows that the
two-parameter family of solutions (\ref{ape07au01}) describe the
tachyon vacuum.

\section{Simple analytic solution in the superstring case}
In this section, we extend our previous results in order to
analyze a new identity-based solution in the modified cubic
superstring field theory, and as in the case of open bosonic
string field theory, by performing a suitable gauge transformation
over this identity-based solution, we obtain the known Gorbachev's
solution which correctly reproduces the value of the vacuum
energy. A two-parameter family of solutions is discussed as well.

\subsection{The identity-based solution and gauge transformation}
In the superstring case, in addition to the basic string fields
$K$, $B$ and $c$, we need to include the super-reparametrization
ghost field $\gamma$ which, in the operator representation, is
given by \cite{Erler:2007xt}
\begin{eqnarray}
\label{gamma01} \gamma &\rightarrow&  U_{1}^\dag U_{1} \tilde
\gamma(0)|0\rangle \, .
\end{eqnarray}
Let us remember that in the superstring case the basic string
fields $K$, $B$, $c$ and $\gamma$ satisfy the algebraic relations
\cite{Erler:2007xt,Gorbachev:2010zz}
\begin{align}
&\{B,c\}=1\, , \;\;\;\;\;\;\; [B,K]=0 \, , \;\;\;\;\;\;\;
B^2=c^2=0
\, , \nonumber\\
\label{02eq2} \partial c = [K&,c] \, , \;\;\;\;\;\;\;
\partial \gamma  = [K,\gamma] \, , \;\;\;\;\;\;\; [c,\gamma]=0 \, ,
\;\;\;\;\;\;\; [B,\gamma]=0 \, ,
\end{align}
and have the following BRST variations
\begin{eqnarray}
\label{02eq3} QK=0 \, , \;\;\;\;\;\; QB=K \, , \;\;\;\;\;\;
Qc=cKc-\gamma^2 \, , \;\;\;\;\;\; Q\gamma=c \partial \gamma
-\frac{1}{2} \gamma
\partial c \, .
\end{eqnarray}
Employing these basic string fields, we can construct the
following identity-based solution
\begin{eqnarray}
\label{02eq1} \Psi=(c+B \gamma^2)(1-K)
\end{eqnarray}
which formally satisfies the equation of motion $Q \Psi + \Psi^2
=0$, where in this case $Q$ is the BRST operator of the open
Neveu-Schwarz superstring theory.

As in the bosonic case, the direct evaluation of the vacuum energy
using the identity-based solution (\ref{02eq1}) brings ambiguous
result, therefore we should try to find an explicit gauge
transformation in order to construct a well behaved solution
similar to Erler-Schnabl's solution (\ref{ESsolution}). Using the
same procedure developed in the previous section, we show that a
well behaved solution $\Psi_{\text{G}}$ can be generated from our
identity-based solution (\ref{02eq1}) by performing a gauge
transformation
\begin{eqnarray}
\Psi &=& U^{-1} (Q + \Psi_{\text{G}}) U \nonumber \\
&=& \Big[1-cBK\frac{1}{1+K}\Big] \Big( Q + (c +
cKBc+B\gamma^2)\frac{1}{1+K}
\Big) \Big[cBK +1\Big] \nonumber \\
&=& \Big[1-cBK\frac{1}{1+K}\Big] \Big( Q(cKB+1) +
(c+cKBc+B \gamma^2)\frac{1}{1+K}(cBK+1) \Big) \nonumber \\
&=& \Big[1-cBK\frac{1}{1+K}\Big] \Big( cKcKB -cK^2 + c(1-K) + cKBc + B \gamma^2 (1-K) \Big) \nonumber \\
&=& \Big[1-cBK\frac{1}{1+K}\Big] \Big( (cBK+1)c(1-K) + B \gamma^2 (1-K) \Big) \nonumber \\
&=&\Big[1-cBK\frac{1}{1+K}\Big] (cBK+1) \Big( (c+B \gamma^2)(1-K)
\Big)
\nonumber \\
&=& (c+B \gamma^2)(1-K) \, .
\end{eqnarray}
Remarkably, it turns out that the resulting analytic solution
$\Psi_{\text{G}}$ corresponds to the known Gorbachev's solution
\cite{Gorbachev:2010zz}
\begin{eqnarray}
\label{Solsimple} \Psi_{\text{G}} &=& (c +
cKBc+B\gamma^2)\frac{1}{1+K} \,
\end{eqnarray}
which, as we are going to verify in the next subsection, correctly
reproduces the value of the vacuum energy.

Since the solution (\ref{Solsimple}) looks very similar to the
bosonic one (\ref{ESsolution}), we should write it as a continuous
integral over wedge states by inverting $1+K$ and using the
Schwinger parameterization (\ref{1K01}), this way of writing the
solution is important if we are interesting in the evaluation of
the vacuum energy. Let us mention that the calculation of the
vacuum energy for the solution (\ref{Solsimple}) was already
performed in reference \cite{Gorbachev:2010zz} by Gorbachev.
Nevertheless, for completeness reasons, we are going to review
this important computation in the next subsection.

\subsection{Gorbachev's solution and its vacuum energy}
As it was mentioned in the previous subsection, by inverting $1+K$
and employing the Schwinger parameterization (\ref{1K01}), we can
write the solution (\ref{Solsimple}) as the following integral
\begin{eqnarray}
\label{ISuperol01} \Psi_{\text{G}} = \int_{0}^{\infty} dt \,
e^{-t}(c + cKBc+B\gamma^2) \Omega^t \, ,
\end{eqnarray}
now acting the BRST operator $Q$ on the string field $Bc$:
$Q(Bc)=cKBc+B \gamma^2$, we can express (\ref{ISuperol01}) as
\begin{eqnarray}
\label{ISuperol02} \Psi_{\text{G}} = \int_{0}^{\infty} dt \,
e^{-t}\, c \, \Omega^t + Q \big[ \int_{0}^{\infty} dt \, e^{-t}\,
Bc \, \Omega^t \big]\, .
\end{eqnarray}

This way of writing the solution (\ref{ISuperol02}) is important
to simplify the computation of the vacuum energy. Using the
equation of motion, the expression for the vacuum energy can be
reduced to the evaluation of the following correlator
\begin{eqnarray}
\label{vacuum02} \frac{1}{6 g^2} \langle  \langle \Psi_{\text{G}}
Q \Psi_{\text{G}} \rangle \rangle&=& \frac{1}{6 g^2}
\int_{0}^{\infty} dt_1dt_2\, e^{-t_1-t_2} \langle \langle c
\Omega^{t_1} (cKc-\gamma^2) \Omega^{t_2}
 \rangle \rangle \nonumber \\
&=& \frac{1}{6 g^2} \int_{0}^{\infty} dt_1dt_2\, e^{-t_1-t_2}
\big[ \langle Y_{-2} c \Omega^{t_1} cKc \Omega^{t_2}
 \rangle - \langle Y_{-2} c \Omega^{t_1} \gamma^2 \Omega^{t_2}
 \rangle \big] \nonumber \\
 &=& -\frac{1}{6 g^2} \int_{0}^{\infty} dt_1dt_2\, e^{-t_1-t_2}
\langle Y_{-2} c \Omega^{t_1} \gamma^2 \Omega^{t_2}
 \rangle  \, ,
\end{eqnarray}
the correlator $\langle Y_{-2} c \Omega^{t_1} \gamma^2
\Omega^{t_2} \rangle $ can be computed using the methods given in
\cite{Erler:2007xt} \footnote{This correlation function has been
computed using the normalization: $\langle \xi(x) c \partial c
\partial^2 c (y) e^{-2 \phi(z)} \rangle=2.$}
\begin{eqnarray}
\label{corre02} \langle Y_{-2} c \Omega^{t_1} \gamma^2
\Omega^{t_2} \rangle = \frac{(t_1+t_2)^2}{2 \pi^2} \, ,
\end{eqnarray}
plugging the value of the correlator (\ref{corre02}) into the
equation (\ref{vacuum02}), we finally obtain
\begin{eqnarray}
\label{vacuum03} \frac{1}{6 g^2} \langle  \langle \Psi_{\text{G}}
Q \Psi_{\text{G}} \rangle \rangle
 &=& -\frac{1}{12 \pi^2 g^2} \int_{0}^{\infty} dt_1dt_2\, e^{-t_1-t_2}
(t_1+t_2)^2  \nonumber \\
&=&-\frac{1}{12 \pi^2 g^2} \big( \int_{0}^{\infty} du \, u^3
e^{-u}\big) \big( \int_{0}^{1} dv \big) \nonumber \\
&=& -\frac{1}{2 \pi^2 g^2} \, ,
\end{eqnarray}
where we have used the change of variables defined as follows
\cite{Erler:2009uj}
\begin{eqnarray}
\label{variables1} u&=& t_1+t_2 \, , \;\;  u \in [0,\infty),
\nonumber \\
v&=& \frac{t_1}{t_1+t_2} \, , \;\;  v \in [0,1], \nonumber \\
dt_1dt_2&=&ududv \, .
\end{eqnarray}

Note that the value of the vacuum energy obtained in equation
(\ref{vacuum03}) is in perfect agreement with the value predicted
from Sen's first conjecture (\ref{poten1}). We would like to
comment that the value of the vacuum energy can be obtained using
another means, for instance using the $\mathcal{L}_{0}$-level
expansion of the solution, we have confirmed that the value of the
vacuum energy is the same as the one computed analytically
(\ref{vacuum03}). It should be important to confirm this result by
using a third option, namely by employing the usual Virasoro
$L_0$-level expansion.

\subsection{Two-parameter family of solutions in the superstring case}
Following the same procedures developed in subsection 2.3, in the
case of the modified cubic superstring field theory, in the
appendix section, we show that Gorbachev's solution
(\ref{Solsimple}) appears as a particular case of a rather general
two-parameter family of solutions
\begin{align}
\label{ape07supersoll} \Psi' =& \big[ x_1 c + x_2 cK +x_3 Kc +x_4
B
\gamma^2 + x_5 B \gamma^2 K +x_6 KB \gamma^2 \big] \frac{1}{1+ K} \nonumber \\
&+ Q \Big\{ \big[ \beta_1 Bc + (\beta_1 + \beta_2) BcK - (1 +
\beta_2) KBc \big] \frac{1}{1+K} \Big\} \, ,
\end{align}
where the coefficients $x_i$ are given by
\begin{align}
\label{xxx123soll} x_1 &= \beta_1 \, , \;\;\;\; x_2= \beta_1 +
\beta_2 \, , \;\;\;\; x_3=-1 - \beta_2 \, , \;\;\; x_4 = 1 -
\beta_1 \, , \nonumber
\\ x_5&=-\frac{(\beta _1-1) (\beta _1+\beta _2)}{\beta _1} \, ,
\;\;\; x_6=\frac{(\beta _1-1) (\beta _2+1)}{\beta _1} \, .
\end{align}
Notice that Gorbachev's solution corresponds to the particular
values of the parameters $\beta_1=1$ and $\beta_2=-1$.
Nevertheless it is interesting to ask: do the solutions with
different values of the parameters describe the tachyon vacuum? To
provide an answer to this question, we should compute the vacuum
energy for the solution (\ref{ape07supersoll}) with arbitrary
values for the parameters $\beta_1$ and $\beta_2$.

Since the solution (\ref{ape07supersoll}) has an expression
containing an exact BRST term, the computation of the vacuum
energy can be reduced to the evaluation of the following
correlator
\begin{align}
\label{correxz02super} \frac{1}{6 g^2}\langle \langle \Psi',Q
\Psi' \rangle \rangle = \frac{1}{6 g^2} \langle \langle \Psi_1,Q
\Psi_1 \rangle \rangle\, ,
\end{align}
where the string field $\Psi_1$ is given by
\begin{align}
\label{ape07supersol2} \Psi_1 = \big[ x_1 c + x_2 cK +x_3 Kc +x_4
B \gamma^2 + x_5 B \gamma^2 K +x_6 KB \gamma^2 \big] \frac{1}{1+
K} \, .
\end{align}
After a tedious algebra, by plugging the string field defined in
(\ref{ape07supersol2}) into (\ref{correxz02super}), we obtain the
following expression for the vacuum energy
\begin{align}
\label{correxz04super} &\frac{1}{6 g^2} \big[ -\langle \langle
c\frac{1}{K+1}\gamma ^2\frac{1}{K+1}\rangle \rangle x_1^2 -2
\langle \langle cK\frac{1}{K+1}\gamma ^2\frac{1}{K+1}\rangle
\rangle x_1 x_2 -2 \langle \langle c\frac{1}{K+1}K\gamma ^2\frac{1}{K+1}\rangle \rangle x_1 x_3  \nonumber \\
&-2 \langle \langle cK\frac{1}{K+1}K\gamma ^2\frac{1}{K+1}\rangle
\rangle x_2 x_3 -\langle \langle cK\frac{1}{K+1}\gamma
^2K\frac{1}{K+1}\rangle \rangle  x_2^2 -\langle \langle
Kc\frac{1}{K+1}K\gamma ^2\frac{1}{K+1}\rangle \rangle x_3^2
\nonumber \\
&+4 \langle \langle B\gamma ^2\frac{1}{K+1}cKc\frac{1}{K+1}\rangle
\rangle x_1 x_4+ 3 \langle \langle B\gamma
^2K\frac{1}{K+1}cKc\frac{1}{K+1}\rangle \rangle x_1 x_5 \nonumber \\
&+ \langle \langle B\gamma ^2\frac{1}{K+1}cKcK\frac{1}{K+1}\rangle
\rangle x_2 x_4  +  3 \langle \langle B\gamma
^2\frac{1}{K+1}KcKc\frac{1}{K+1}\rangle \rangle x_3 x_4 \nonumber \\
&+\langle \langle KB\gamma ^2\frac{1}{K+1}cKc\frac{1}{K+1}\rangle
\rangle x_1 x_6 \big].
\end{align}
All the correlators involved in the evaluation of the vacuum
energy (\ref{correxz04super}) can be computed using the basic
correlators
\begin{eqnarray}
\langle\langle \Omega^{r_1}c \Omega^{r_2} \gamma^2 \Omega^{r_3}
\rangle\rangle
&=&\frac{(r_1+r_2+r_3)^2}{2 \pi ^2} \, ,\\
\langle\langle\Omega^{r_1} B \Omega^{r_2}
c\Omega^{r_3}c\Omega^{r_4} \gamma^2\Omega^{r_5} \rangle\rangle &=&
\frac{(r_1+r_2+r_3+r_4+r_5) r_3 }{2 \pi ^2} \, .
\end{eqnarray}
For instance, the expression for the correlator $\langle \langle
B\gamma ^2\frac{1}{K+1}cKc\frac{1}{K+1}\rangle \rangle$ is given
by
\begin{align}
\label{correxz06super} \langle \langle B\gamma
^2\frac{1}{K+1}cKc\frac{1}{K+1}\rangle \rangle &= -
\int_{0}^{\infty}\int_{0}^{\infty}dt_1dt_2 e^{-t_1-t_2}
\frac{\partial \langle\langle B \gamma^2 \Omega^{t_1} c
\Omega^{s_1}  c \Omega^{t_2} \rangle\rangle}{\partial s_1}
\Big|_{s_1=0} \nonumber \\
&= - \frac{1}{2 \pi^2} \int_{0}^{\infty}\int_{0}^{\infty}dt_1dt_2
e^{-t_1-t_2} (t_1+t_2) \nonumber\\
&=-\frac{1}{\pi ^2} \, .
\end{align}

Performing similar computations for the rest of the correlators,
from (\ref{correxz04super}) we get the following expression for
the vacuum energy (\ref{correxz02super})
\begin{align}
\label{correxz07super} \frac{1}{6 g^2}\langle \langle  \Psi',Q
\Psi' \rangle \rangle =\frac{-6 x_1^2+\left(8 x_2+8 x_3-4 x_4+3
x_5+x_6\right) x_1-2 \left(x_2+x_3\right)^2+\left(x_2+3 x_3\right)
x_4}{12 g^2 \pi
   ^2}
\, ,
\end{align}
plugging the definition of the coefficients $x_1$, $x_2$, $x_3$,
$x_4$, $x_5$, $x_6$ (\ref{xxx123soll}) into the equation
(\ref{correxz07super}), we remarkably obtain the right value for
the vacuum energy
\begin{align}
\label{correxz08super} \frac{1}{6 g^2}\langle \langle  \Psi',Q
\Psi' \rangle \rangle = -\frac{1}{2 \pi ^2 g^2 } \, ,
\end{align}
therefore this last result (\ref{correxz08super}) shows that the
two-parameter family of solutions (\ref{ape07supersoll}) describe
the tachyon vacuum.

\section{Summary and discussion}

We have studied a new set of identity-based solutions to analyze
the problem of tachyon condensation in open bosonic string field
theory and cubic superstring field theory, although these
solutions seem to be trivial, we have shown that they can be
related, by performing a gauge transformation, to well behaved
solutions where in the case of open bosonic string field theory,
the resulting solution corresponds to the Erler-Schnabl's solution
\cite{Erler:2009uj}, while in the case of the modified cubic
superstring field theory corresponds to a similar solution
\cite{Gorbachev:2010zz} which, unlike the known solutions
\cite{Erler:2007xt,Aref'eva:2009sj,Aref'eva:2008ad}, can be
written as a continuous integral over wedge states where no
regularization or phantom term is required.

Although, after performing the gauge transformation, the resulting
Erler-Schnabl-type solutions can be used to compute correctly the
value of the vacuum energy, it would be interesting to evaluate
directly the vacuum energy using the identity-based solutions,
this kind of computations should be possible provide that we can
find a consistent regularization scheme. Even though our results
suggest that these solutions should reproduce the right value for
the D-brane tension, the direct calculation persists as one of the
most difficult problems in string field theory
\cite{Kishimoto:2009nd}.

The main motivation of this work was to understand how
identity-based solutions can be used to generate well defined
solutions which describe to the tachyon vacuum in relatively
simple cubic string field theories
\cite{Preitschopf:1989fc,Arefeva:1989cp,Witten:1985cc}. It would
be important to extend this analysis to the case of Berkovits
WZW-type superstring field theory \cite{Berkovits:1995ab}, since
this theory has a non-polynomial action, the issue for finding the
tachyon vacuum solution and the computation of the value of the
D-brane tension seems to be highly cumbersome. Nevertheless, we
hope that the ideas developed in this paper should be very useful
in order to solve this challenging puzzle.

Another important application of the techniques developed in this
paper, as mentioned in \cite{Zeze:2010jv}, should be the extension
of the subalgebra generated by the basic string fields $K$, $B$,
$c$ and $\gamma$ in order to analyze more general string field
configurations \cite{Garousi:2008ge,Garousi:2007fk}, such that
multiple D-branes, marginal deformations, lump solutions as well
as time dependent solutions.

\section*{Acknowledgements}
I would like to thank Nathan Berkovits, Ted Erler, Michael Kroyter
and Martin Schnabl for useful discussions. This work is supported
by CNPq grant 150051/2009-3.

\appendix
\setcounter{equation}{0}

\def\thesection{\Alph{section}}
\renewcommand{\theequation}{\Alph{section}.\arabic{equation}}

\section{Derivation of the identity-based
solution and gauge transformation} In this appendix we provide the
details related to the derivation of the identity-based solution
and the gauge transformation. In the case of open bosonic string
field theory, we propose the following ansatz
\begin{eqnarray}
\label{ansa} \Psi = \alpha_1 c + \alpha_2 cK + \alpha_3 Kc \, ,
\end{eqnarray}
plugging this ansatz into the equation of motion $Q \Psi + \Psi
\Psi=0$ and using the identity $cK^2c=cKcK+KcKc$, we find easily
\begin{eqnarray}
\label{ape01} \alpha_1(1+\alpha_2+\alpha_3) cKc +
\alpha_2(1+\alpha_2+\alpha_3) cKcK + \alpha_3(1+\alpha_2+\alpha_3)
KcKc=0 \, .
\end{eqnarray}
Therefore, as we can see, the string field equation of motion
reduces to the algebraic equation
\begin{eqnarray}
\label{ape02} 1+\alpha_2+\alpha_3 =0 \, ,
\end{eqnarray}
and consequently our ansatz (\ref{ansa}) becomes
\begin{eqnarray}
\label{ape03} \Psi = \alpha_1 c + \alpha_2 cK - (1+\alpha_2) Kc \,
.
\end{eqnarray}

The next step is to perform a gauge transformation over this
identity-based solution (\ref{ape03}). It turns out that the
suitable string field $U$ which will be used to define the gauge
transformation is given by
\begin{eqnarray}
\label{ape04} U= 1 + cBK \, .
\end{eqnarray}
The inverse of this string field $U$ can be computed using the
power series expansion
\begin{eqnarray}
\label{ape05} U^{-1}= \frac{1}{1 + cBK} = \sum_{n=0}^{\infty}
(-1)^n (cBK)^n = 1+\sum_{n=1}^{\infty} (-1)^n cB K^n = 1- cBK
\frac{1}{1+K} \, .
\end{eqnarray}

A string field $\Psi'$ which identically satisfies the string
field equation of motion $Q \Psi' + \Psi' \Psi'=0$ can be derived
by performing a gauge transformation over the identity-based
solution (\ref{ape03})
\begin{eqnarray}
\label{ape06} \Psi'=U (\Psi+Q) U^{-1} \, .
\end{eqnarray}
Plugging the expressions (\ref{ape04}) and (\ref{ape05}) for the
string field $U$ and its inverse $U^{-1}$ into the definition of
the gauge transformation (\ref{ape06}), we obtain
\begin{eqnarray}
\label{ape07} \Psi'= \big[ \alpha_1 (c+ cBKc) + (\alpha_1 +
\alpha_2)( cK + cBKcK) -(1+\alpha_2)( Kc + KcBKc )  \big]
\frac{1}{1+K} \, . \nonumber \\
\end{eqnarray}

To simplify the computation of the vacuum energy, we would like to
write Erler-Schnabl-type solution (\ref{ape07}) in the following
way
\begin{eqnarray}
\label{ape08} \Psi'= \big[ \alpha_1 c + \chi  \big] \frac{1}{1+K}
\end{eqnarray}
such that $Q \chi=0$ where $\chi$ is some string field, to satisfy
this requirement (\ref{ape08}) we must impose the following
condition on the numerator of equation (\ref{ape07})
\begin{eqnarray}
\label{ape09} Q \big[ \alpha_1  (c+cBKc) + (\alpha_1 + \alpha_2)(
cK + cBKcK) -(1+\alpha_2)( Kc + KcBKc )  \big] = \alpha_1  cKc  \nonumber \\
\end{eqnarray}
from this last equation (\ref{ape09}), using the BRST variations
(\ref{eq3}), we obtain
\begin{eqnarray}
\label{ape10} (\alpha_1 + \alpha_2)cKcK  -(1+\alpha_2)KcKc = 0 \,
,
\end{eqnarray}
and therefore the value of the coefficients $\alpha_1$ and
$\alpha_2$ are given by
\begin{eqnarray}
\label{ape11} \alpha_1 =1 \;\;\; , \;\;\;   \alpha_2=-1 \, .
\end{eqnarray}

Finally, plugging the value of these coefficients (\ref{ape11})
into Erler-Schnabl-type solution (\ref{ape07}), we get the well
known Erler-Schnabl's tachyon vacuum solution in open bosonic
string field theory \cite{Erler:2009uj}
\begin{eqnarray}
\label{ape12} \Psi_{\text{E-S}}= \big[ c+ cBKc \big] \frac{1}{1+K}
\, .
\end{eqnarray}

In the case of the modified cubic superstring field theory, we
should use the following ansatz
\begin{align}
\label{ansasuper01} \Psi=\sum_{n,p} f_{n,p}U_{1}^\dag U_{1}
\hat{\mathcal{L}}^n \tilde{c}_p |0\rangle +\sum_{n,p,q} f_{n,p,q}
U_{1}^\dag U_{1} \hat{\mathcal{B}}\hat{\mathcal{L}}^n \tilde{c}_p
\tilde{c}_q |0\rangle + \sum_{n,t,u} g_{n,t,u} U_{1}^\dag U_{1}
\hat{\mathcal{B}}\hat{\mathcal{L}}^n \tilde{\gamma}_t
\tilde{\gamma}_u |0\rangle \, ,
\end{align}
where $n = 0, 1, 2, \cdots$, $p, q = 1, 0,-1,-2, \cdots $ and $t,
u = \frac{1}{2},-\frac{1}{2},-\frac{3}{2}, \cdots \cdot$ Plugging
this ansatz (\ref{ansasuper01}) into the equation of motion will
lead to a system of algebraic equations for the coefficients
$f_{n,p}$, $f_{n,p,q}$ and $g_{n,t,u}$. Analyzing these algebraic
equations we discover that many of the coefficients can be set to
zero, therefore we can use a simpler ansatz than the one given by
(\ref{ansasuper01}), namely
\begin{eqnarray}
\label{ansa002super} \Psi = \beta_1 c + \beta_2 cK + \beta_3 Kc +
\beta_4 B\gamma^2 +  \beta_5 B\gamma^2K + \beta_6 KB\gamma^2 \, .
\end{eqnarray}
Plugging this ansatz (\ref{ansa002super}) into the equation of
motion $Q \Psi + \Psi \Psi=0$, we obtain a system of algebraic
equations for the coefficients $\beta_i$
\begin{eqnarray}
\label{parametersuper01} \beta _2+\beta _3+1&=&0  \\
-\beta _2 \beta _5+\beta _3 \beta _5-\beta _5-\beta _2 \beta
_6+\beta _3
   \beta _6+\beta _6&=&0   \\ \beta _2-\beta _3-\beta _1 \beta _5+\beta _1 \beta
   _6&=&0  \\ \beta _4-1&=&0 \, .
\end{eqnarray}
Solving this system of algebraic equations, we obtain
\begin{eqnarray}
\label{parametersuper02} \beta_3= -\beta _2-1 \, , \;\;\;\;
\beta_4=1 \, , \;\;\;\; \beta_5=\frac{\beta _2}{\beta _1} \, ,
\;\;\;\; \beta_6=\frac{-\beta _2-1}{\beta _1} \, ,
\end{eqnarray}
and consequently our ansatz (\ref{ansa002super}) becomes
\begin{eqnarray}
\label{ape03super}\Psi = \beta_1 c + \beta_2 cK -(\beta _2+1) Kc +
 B\gamma^2 +  \frac{\beta _2}{\beta _1} B\gamma^2K - \frac{\beta _2+1}{\beta _1} KB\gamma^2 \, .
\end{eqnarray}

A string field $\Psi'$ which identically satisfies the string
field equation of motion can be derived, as in the bosonic case,
by performing a gauge transformation over the identity-based
solution (\ref{ape03super})
\begin{eqnarray}
\label{ape06super} \Psi'=U (\Psi+Q) U^{-1} \, .
\end{eqnarray}
Plugging the expressions (\ref{ape04}) and (\ref{ape05}) for the
string field $U$ and its inverse $U^{-1}$ into the definition of
the gauge transformation (\ref{ape06super}), we obtain a
two-parameter family of solutions
\begin{align}
\label{ape07super} \Psi' =& \big[ x_1 c + x_2 cK +x_3 Kc +x_4 B
\gamma^2 + x_5 B \gamma^2 K +x_6 KB \gamma^2 \big] \frac{1}{1+ K} \nonumber \\
&+ Q \Big\{ \big[ \beta_1 Bc + (\beta_1 + \beta_2) BcK - (1 +
\beta_2) KBc \big] \frac{1}{1+K} \Big\} \, ,
\end{align}
where the coefficients $x_i$ are given by
\begin{align}
\label{xxx123} x_1 &= \beta_1 \, , \;\;\;\; x_2= \beta_1 + \beta_2
\, , \;\;\;\; x_3=-1 - \beta_2 \, , \;\;\; x_4 = 1 - \beta_1 \, ,
\nonumber
\\ x_5&=-\frac{(\beta _1-1) (\beta _1+\beta _2)}{\beta _1} \, ,
\;\;\; x_6=\frac{(\beta _1-1) (\beta _2+1)}{\beta _1} \, .
\end{align}
These coefficients have been defined to simplify the presentation
of our equations. Notice that the Gorbachev's solution
\cite{Gorbachev:2010zz}
\begin{eqnarray}
\label{ape13} \Psi_{\text{G}} = (c + cKBc+B\gamma^2)\frac{1}{1+K}
\end{eqnarray}
corresponds to the particular case $\beta_1=1$ and $\beta_2=-1$.

\newpage

\end{document}